\documentstyle[12pt]{article}

\renewcommand{\ref}[1]{\raisebox{.6ex}{[#1]}}

\newcommand{\be}{\begin{equation}}
\newcommand{\ee}{\end{equation}}

\newcommand{\ba}{\begin{array}}
\newcommand{\ea}{\end{array}}

\begin{document}



\title{ Binary Bose-Einstein Condensate Mixtures in Weakly and Strongly 
        Segregated Phases  }

\author{ P. Ao$^{\star}$ and S.T. Chui    \\
          Bartol Research Institute \\
    University of Delaware, Newark, DE 19716  }

\maketitle


\begin{abstract}
We perform a mean-field study of the binary Bose-Einstein
condensate mixtures as a function of the mutual repulsive 
interaction strength. 
In the phase segregated regime, we find that there are two
distinct phases: the weakly segregated phase characterized 
by a `penetration depth' and the strongly segregated phase 
characterized by a healing length.
In the weakly segregated phase the symmetry of the shape
of each condensate will not take that of 
the trap because of the finite surface tension, 
but its total density profile still does.
In the strongly segregated phase even the total density profile
takes a different symmetry from that of the trap
because of the mutual exclusion of the condensates.
The lower critical condensate-atom number to observe the
complete phase segregation is discussed.
A comparison to recent experimental data suggests that the weakly 
segregated phase has been observed.

\noindent
PACS${\#}$:  03.75.Fi
 
\end{abstract}

 

\section{Introduction}

Shortly after its first theoretical study \cite{ho}, the binary mixture of 
Bose-Einstein condensates (BEC's) of alkali-metal atoms in a trap
has been realized experimentally \cite{jila},
and the development of this field is now 
blooming \cite{further,molmer,theoretical,law,cta}.
The binary mixture idea has been extended to trapped boson-fermion
and fermion-fermion systems \cite{molmer}.
This opens the door to studying the rich physics in new parameter regimes.
There have been several theoretical studies \cite{theoretical,law,cta};
most of them are numerical in nature
or are from the atomic physics point of view.
In the present paper, we perform a mean-field-type study based on
the nonlinear Schr\"odinger equation, and obtain
a qualitative and in many cases a quantitative analytical understanding
of a variety of properties of the binary BEC mixtures.
Along with results from 
the competition between the healing length 
and the `penetration depth', as well as from the finite trap effect, 
we have summarized previously known results in the manner of simplified 
mean-field solutions. 
In this way, we provide a convenient framework for
classifying various excitations in the system, 
and pave the way for further study of properties of binary BEC's, such
as the time evolution of the two condensates 
during a phase segregating process.	

We start from the Hamiltonian formulation of the binary BEC's at zero 
temperature:
\[
   H = \int d^3 x \left[ \psi^\ast_1(x) 
       \left( \frac{ - \hbar^2\nabla^2 }{2m_1} \right) \psi_1(x) 
       + \psi^\ast_1(x) U_1(x) \psi_1(x) \right]
\]
\[  
       +  \int d^3 x \left[ \psi^\ast_2(x) 
       \left( \frac{ - \hbar^2\nabla^2 }{2m_2} \right) \psi_2(x) 
       + \psi^\ast_2(x) U_2(x) \psi_2(x) \right]  \\
\]
\[
    + \frac{G_{11}}{2} \int d^3 x \; \psi^\ast_1(x) \psi_1(x) \,
                                 \psi^\ast_1(x) \psi_1(x)  
\]
\[
    + \frac{G_{22}}{2} \int d^3 x \; \psi^\ast_2(x) \psi_2(x) \,
                                 \psi^\ast_2(x) \psi_2(x) 
\]  
\be
    + G_{12} \int d^3 x \; \psi^\ast_1(x) \psi_1(x) \,
                                 \psi^\ast_2(x) \psi_2(x)  \; .
\ee  
Here $\psi_i$, with $i=1,2$, 
is the effective wave function of the $i$th 
condensate, with the mass $m_i$ and the trapping potential $U_i$.
The interaction between the $i$th condensate atoms is specified by $G_{ii}$,
and that between 1 and 2  by $G_{12}$.
In the present paper all $G$'s will be taken to be positive.
The corresponding time independent equation of motion 
is the well-known  nonlinear Schr\"odinger equation \cite{nlse},
obtained here by minimization of the energy, Eq. (1), 
with fixed condensate atom numbers:
\be
  -\frac{\hbar^2}{2m_1} \nabla^2 \psi_1(x)  + (U_1(x) - \mu_1 ) \psi_1(x)
    + G_{11} |\psi_1(x)|^2 \psi_1(x) 
    + G_{12} |\psi_2(x)|^2 \psi_1(x) = 0 \; ,
\ee
\be
  -\frac{\hbar^2}{2m_2} \nabla^2 \psi_2(x)  + (U_2(x) - \mu_2 ) \psi_2(x)
    + G_{22} |\psi_2(x)|^2 \psi_2(x) 
    + G_{12} |\psi_1(x)|^2 \psi_2(x) = 0  \; .
\ee
  The Lagrangian multipliers, the chemical potentials $\mu_1$ and $\mu_2$, 
are determined by the relations
$ \int d^3 x  |\psi_i(x)|^2  =  N_i \; , i =1,2 $,
with $N_i$ the number of the $i$th condensate atoms.

Experimentally, the trapping potentials $\{ U_i \}$ are
simple harmonic in nature.
For the sake of simplicity and to illustrate the physics
we shall consider a square well tapping potential $U_i = U$:
zero inside and large (infinite) outside,
unless otherwise explicitly specified.

\section{Criteria and Symmetries in Segregated Phases }

\subsection{ Simplified mean-field solutions}

With the square well trapping potential specified in Sec. 1,
 the coupled nonlinear Schr\"odinger equations
have an obvious homogeneous solution:
inside the trap the condensate densities $\rho_i = |\psi_i|^2$,
$
   \rho_i =  \frac{N_i}{V} \; , 
$
with $V$ the volume of the square well potential trap,
and the chemical potentials
$
   \mu_1 = G_{11} \rho_1 + G_{12}\rho_2   
$
and
$ 
   \mu_2 = G_{22} \rho_2 + G_{12}\rho_1   
$.
The corresponding total energy of the system is
\be
   E_{ho} = \frac{1}{2} \left[ G_{11} \frac{N_1^2}{V} + 
       G_{22} \frac{N_2^2}{V} + 2 G_{12} \frac{N_1 N_2 }{V} \right]  \; .
\ee
For a small enough $G_{12}$, any variation on top of this solution 
will increase the system energy. 
This implies that the excitations are stable.
Therefore in this parameter regime the homogeneous state is the ground state.
If the mutual repulsive interaction $G_{12}$  is strong enough, however,
this is no longer true. We show below in a mean-field manner that 
there is an inhomogeneous solution with a lower 
total system energy.

Let us consider the case of the inhomogeneous state
in which the two  condensates mutually exclude each other.
For the moment we ignore the thickness of the interface and the 
corresponding extra energy. In this way we temporarily ignore
the derivative terms in Eqs. (2) and (3) 
in determining the effective condensate wave functions.
We call this situation the simplified mean-field approach, which is a useful
one that has already given us a lot of physical insights \cite{ho}.
Let $V_i$ be the volume inside the trap occupied by the condensate $i$.  
We have 
$|\psi_i |^2 = \rho_{i0} =  N_i/V_i$
and the total energy of the inhomogeneous state
$ E_{in} = \frac{1}{2}\sum_{i=1,2} G_{ii} \frac{N_i^2}{V_i} $.
Minimizing $E_{in}$ with respect to $V_1$ or $V_2$ 
under the constraint
$V_1 + V_2 = V$, 
 we obtain the spatial volume occupied by each condensate:
\[
  V_1 = \frac{1}{ 1+\sqrt{\frac{G_{22}}{G_{11}} }\frac{N_2}{N_1} } V \; , \;
  V_2 = \frac{1}{ 1+\sqrt{\frac{G_{11}}{G_{22}} }\frac{N_1}{N_2} } V  \; .
\]
The corresponding condensate densities are 
\be
  \rho_{10} = \left( 1 + \sqrt{\frac{ G_{22} }{ G_{11} } } 
                 \frac{N_2}{N_1}\right)
                \frac{N_1}{V} \; , \; 
  \rho_{20} = \sqrt{ \frac{G_{11}}{G_{22} } }  \; \rho_{10} \; ,
\ee
and the chemical potentials 
$
  \mu_i = G_{ii} \; \rho_{i0}  
$.
We note here $\mu_1 \rho_{10} = \mu_2 \rho_{20} $. 
The total energy for this inhomogeneous state is 
\[
    E_{in} = \frac{1}{2} \left[
         G_{11} \frac{N_1^2}{V } 
       +  G_{22} \frac{N_2^2}{V } 
       +  2\sqrt{G_{11}G_{22} } \frac{N_1 N_2 }{V } \right]   \; .
\]
The energy difference from the homogeneous state is then
\be
   \Delta E = E_{in} - E_{ho} = 
   - \left( G_{12} - \sqrt{G_{11}G_{22} }  \right) 
      \frac{N_1 N_2}{V}  \; .
\ee
This equation reveals that 
for a large enough mutual repulsive interaction, 
that is, if
\be
   G_{12} > \sqrt{ G_{11} G_{22} }  \; ,
\ee
the inhomogeneous state has a lower total energy.
Hence, the inhomogeneous state, the phase segregation state, 
will be favored for a large mutual repulsive interaction $G_{12}$.
We note that this criterion for the mutual repulsive interaction strength
is independent of the condensate-atom numbers
as well as of the trap size. We shall return to this point below.
The critical value for $G_{12}$, Eq. (7), has been found using a
stability analysis from the excitation spectrum in
the homogeneous state \cite{law}, 
while it is obtained here from a simple energetic consideration.

\subsection{ Interface profile }

In the absence of the derivative terms in the determination of 
the condensate wave functions studied above, 
the shape of the boundary 
between the two condensates in the phase segregated 
state will take any form. 
Now, we consider the effect of the derivative terms.
Their inclusion  will make the thickness of the interface finite 
and introduce a finite interface energy, the surface tension.
The presence of the surface tension
will fix the shape of the interface between the two condensates
by minimization of the total surface energy.
First, we look for the condensate profiles at the interface with a finite 
thickness. We rescale the effective wave functions
by the values deep inside their own condensates: 
\[
   \psi_i = f_i  \sqrt{\rho_{i 0}}  \; .
\]
We shall assume here that  condensate 1 occupies the region in the
trap specified by $z>0$, and condensate 2 occupies the region 
specified by $z<0$. The interface plane is $z=0$.

Deep inside the region of condensate 1, we have $f_1 \rightarrow 1$
and $ f_2 \rightarrow 0$.
In this region, using $\mu_i = G_{ii} \rho_{i0}$ and only 
keeping the leading contributions,
the coupled nonlinear Schr\"odinger 
equations, Eqs. (2) and (3), becomes 
\be
   - \frac{\hbar^2}{2m_1} \nabla^2 \delta f_1 + 2 G_{11}\rho_{10}  
      \; \delta f_1 + G_{12} \rho_{20} f_2^2  = 0  \; ,
\ee
\be
   - \frac{\hbar^2}{2m_2} \nabla^2 f_2 - G_{22}\rho_{20} f_2 
     + G_{12} \rho_{10}  f_2 = 0  \; .
\ee
Here $ f_1 = 1 + \delta f_1$.
These equations may be written in the following
more suggestive form
\[
   - \xi_1^2 \nabla^2 \delta f_1 + 2  \delta f_1 
    + \frac{G_{12} }{ \sqrt{G_{11} G_{22} } }  f_2^2 
       = 0 \; ,
\]
\[
   - \xi_2^2  \nabla^2 f_2 + \left( 
     \frac{ G_{12} }{ \sqrt{G_{11}G_{22} } } - 1 \right)  f_2 = 0  \; .
\]
with the `healing lengths' $\xi_i$ defined as
\be
  \xi_i  = \sqrt{ \frac{\hbar^2} {2m_i} \frac{1 }{G_{ii} \rho_{i0} }} \; . 
\ee
From Eq. (9) 
we find the density profile of condensate 2 in  condensate 1 is
\be
   f_2(z)  = f_2(0) e^{- \frac{z}{\Lambda_2 } } \; ,
\ee
with the `penetration depth' 
\be
  \Lambda_2 = 
  \frac{1}{ \sqrt{ \frac{G_{12} }{ \sqrt{G_{11} G_{22} } } - 1 } } \xi_2 
       \; ,
\ee
which is the length scale for condensate 2  penetrating into 1.
Similarly, for the penetration of condensate 1 into condensate 2 in
region $z<0$,  we have 
$   f_1(z)  = f_1(0) e^{ {z}/{\Lambda_1 } } $, 
with the penetration depth 
$    \Lambda_1 = 
  {\xi_1}/{ \sqrt{ \frac{G_{12} }{ \sqrt{G_{11} G_{22} } } - 1 } } 
$.
The healing length scale $\xi$ 
here describes the ability of a condensate to 
recover from a disturbance, similar to the same length scale in superfluid
helium 4. 
The newly introduced length scale here, the penetration depth $\Lambda$,
describes the degree of the mixing between the two condensates.
Obviously, as $G_{12} \rightarrow \sqrt{G_{11} G_{22} }$, the 
penetration depth goes to infinity, 
in coincidence with the disappearing of the phase segregation.

It is also useful to study the behavior of  condensate 1 in 
the region $z>0$. 
Deep inside condensate 1, $\delta f_1$ is small.
If $f^2_2$ approaches zero faster than $\delta f_1$, 
that is, $f^2_2 < \delta f_1$,
the last term in Eq. (8) may be dropped, and we have 
\be
  \delta f_1 = \delta f_1(0) \; e^{ - \sqrt{2} z/\xi_1}  \; .
\ee
Here it is the healing length $\xi_1$ of condensate 1, 
not the penetration depth $\Lambda_2$, 
that determines the profile of condensate 1. 
The validity of self-consistency for this solution requires 
$ \Lambda_2 < \sqrt{2} \xi_1$.
In this parameter regime the mutual repulsive interaction is so strong that
condensate 1 stays away from 2, which is similar to the Meissner 
state where the magnetic field is completely
excluded outside of a bulk superconductor.
We shall call this parameter regime the strongly segregated phase.
In the opposite limit, that is, $ \Lambda_2 > \sqrt{2} \xi_1$, 
$\delta f_1$ will be 
determined by $f_2$ through Eq. (8): 
\be
  \delta f_1 = - \frac{1}{2 \left(1-\frac{2\xi_1^2}{\Lambda_2^2 }\right) }
    \frac{G_{12}}{\sqrt{ G_{11} G_{22}}} \; f_2^2  \; , \;
    \Lambda_2 > \sqrt{2}\xi_1  \; . 
\ee 
The only relevant length scale here is the penetration depth,
which is larger than the healing length.
There is still a considerable mixing of the two condensates
in this parameter regime, which we shall call the weakly segregated phase.

\subsection{Surface tension}

With the inclusion of the gradient terms in 
the determination of the condensate wave functions,
the presence of the interface will cost a finite amount of energy.
The surface energy per unit area, the surface tension, may be defined as
$ \sigma = \Delta E_s /S$. Here
$S$ is the interface area.
The energy difference  $\Delta E_s$  may be calculated in the 
following manner:
In the presence of the interface at $z=0$ 
we first solve the full Eqs. (2) and (3) 
with derivative terms for the  condensate
wave functions $\psi_i$, calculate the corresponding total energy 
from Eq. (1), then subtract from this total energy by the amount 
given by  Eq. (5), 
the total energy of the system in the segregated phase without 
the effect of the derivative terms in Eqs. (2) and (3).
Specifically, the energy difference $\Delta E_s$ is 
\[
  \Delta E_s = \int d^3x \left\{ \sum_{i=1,2}  
      \left[ \rho_{i0} f_i \left( \frac{ -\hbar^2 \Delta^2 }{2m_i} 
                           \right)  f_i 
   + \frac{G_{ii}}{2} \rho_{i0}^2 f_i^4    \right] +    
         {G_{12}} \rho_{10}\rho_{20} f_1^2 f_2^2  \right\}
   - \sum_{i=1,2} \frac{G_{ii}}{2} \rho_{i0} N_i  \; .     
\]
Because of the normalization condition 
$  \int d^3 x \; \rho_{i0} f_i^2 = N_i $, 
from Eqs. (2) and (3), 
we obtain the surface tension as
\be
   \sigma = \frac{1}{2} \int_{-\infty}^{\infty} d z 
     \sum_{i=1,2} \rho_{i0}
     f_i(z) \left( - \frac{\hbar^2 \nabla^2 }{2m_i } \right) f_{i}(z) \; .
\ee
Though Eq. (15) is a general expression for the surface tension, 
to gain a concrete understanding, we consider the case that the two 
condensates have an identical set of parameters: $\Lambda_i = \Lambda$ and 
$\xi_i = \xi$. 
In the strongly segregated phase of $\xi  >> \Lambda /\sqrt{2} $,  
Eq. (15) gives 
\be
  \sigma = \frac{\xi}{\sqrt{2} } 
            \; \sqrt{G_{11}G_{22} } \; \rho_{10}\rho_{20} \;  ,
\ee
which is independent of the penetration depth and 
the mutual repulsive interaction.
In the weakly segregated phase of  $ \xi << \Lambda/\sqrt{2} $, 
Eq. (15) gives
\be
  \sigma = \frac{\xi^2}{\Lambda } \; 
           \sqrt{ G_{11} G_{22}} \; \rho_{10} \rho_{20} \; , 
\ee
which goes to zero as $\Lambda \rightarrow \infty$.
We note that this occurs when $G_{12} \rightarrow \sqrt{G_{11} G_{22}}$,
in agreement with our above mean-field analysis of the phase segregation.
The existence of the finite surface tension leads to another branch of 
gapless excitations, the interface or surface mode, which 
we are not going to discuss here.
 
\subsection{ Finite trap size effect I: broken symmetry ground state }

Now we consider the effects of finite surface tension 
and the finite trap size.
For a very large system, it is known that the minimization of surface energy 
leads to the  minimum surface area, whose shape is usually spherical 
 in three dimensions (3D) and circular in 2D.
For a finite size trap, one might expect that the shape of the ground 
state of the binary BEC mixture should take the same symmetry of the
trap, particularly if it is cylindrically or spherically symmetric.   
We show here that this may not be true in a segregated phase, and
the condensates may break the cylindrical symmetry defined by the trap. 
To demonstrate the essential physics, we consider  
the case of two condensates  in two spatial dimensions
with an identical set of parameters:
$\Lambda_i = \Lambda$, $\xi_i = \xi$, and $\rho_{i0} = \rho_0$.  
Let $R$ be the radius of the square 
well potential trap. 
Supposing that condensate 1 occupies a circular area with a radius
$R' = \frac{R}{\sqrt{2}}$; the associated surface energy is
the length of the interface $2\pi R'$  times the surface tension 
$\sigma$ : 
\[
  E_s = \sqrt{2} \; \pi R \sigma \; .
\]
On the other hand, supposing that each condensate occupies
a half circular shape of the trap,
the corresponding surface energy is
\[
  E_b = 2 R \sigma  \; ,
\]
which is much lower than the circular shape with
the same symmetry as the trap,
because the interface length here is shorter than that with
the circular symmetry. 
Therefore, each condensate will take a different shape than
that of the symmetry of the trap. The broken cylindrical symmetry state 
occurs, discovered first numerically \cite{cta}.

Though the circular symmetry is broken for each condensate, 
in the weakly segregated phase where 
$ \Lambda > \sqrt{2} \xi$, 
the total density $\rho_1 + \rho_2$  still
appears circularly symmetric, 
and it retains the symmetry of the trap.
In the strongly segregated phase where 
$\Lambda < \sqrt{2} \xi$, 
the two condensates tend to avoid each other. In this regime
the circular symmetry of the total
density profile is broken. 
Here we give a heuristic demonstration of such different
behaviors of the total density profile in two segregated phases.
We first show that there is a ditch in the total density profile at
the interface in the strongly segregated  phase. 
Following the analysis above, we take $z=0$ as the interface position.
Condensate 1 (2) occupies the $z>0$ $(z<0)$ region  of the trap. 
In the region $z>0$, the condensate densities take the forms
$ \rho_1(z) = \rho_0 ( 1 - b_1 e^{ - \sqrt{2} z/\xi } )$ and
$ \rho_2(z) = \rho_0 \; b_2 e^{ - 2 z/\Lambda } $,
consistent with Eqs. (8) and (9).
Similarly, in the region $z<0$,
$ \rho_1(z) = \rho_0 b_2 e^{ 2 z/\Lambda } $ and
$ \rho_2(z) = \rho_0 ( 1 - b_2 e^{ \sqrt{2} z/\xi } )$.
Here  $b_1$ and $b_2$ 
are two numerical constants.
To determine $b_1$ and $b_2$, we make use of the fact that 
because Eqs. (2) and (3) are second order differential equations,
the solutions and their first order derivatives must be continuous.
This immediately gives us two algebraic equations at the interface $z=0$:
\[
  1 - b_1 = b_2 \; , {\ } 
 \sqrt{2} b_1/\xi = 2 b_2/\Lambda \; .
\]
The numerical constants are then
$  b_1 = \sqrt{2} \xi/(\sqrt{2} \xi + \Lambda) $ and   
$  b_2 = \Lambda / (\sqrt{2}\xi + \Lambda) $.
Evidently, the total density $\rho_1 (z) + \rho_2 (z)$ 
has its minimum value $ \rho_0 \; 2 \Lambda /(\sqrt{2}\xi + \Lambda)$ at
the interface $z=0$, which goes to zero as $\Lambda \rightarrow 0$ or,
equivalently, as $G_{12} \rightarrow \infty$. 
The total density has a ditch at the interface, 
an indication of broken symmetry in the strongly segregated phase.   
For the weakly segregated phase, 
replacing $\sqrt{2}\xi$ by $2\Lambda$ and  following the same procedure 
we obtain $b_1=b_2= 1/2$ 
and find that the total density remains constant in the trap, not affected
by the phase segregation.
There is no broken symmetry for the total density in the weakly
segregated phase.

We note that in terms of interactions the weakly segregated phase is 
specified by  
$ 1 < {G_{12} } / {\sqrt{ G_{11}G_{22} } }  < ( 1 + 1/\sqrt{2} ) $, 
and the strongly segregated phase by 
$  {G_{12} }/{\sqrt{ G_{11}G_{22} } } > ( 1 + 1/\sqrt{2} )$. 

Another interesting feature of the 
finite surface tension is the floating of the condensate droplet.
This may be regarded as a special case of the symmetry broken state.
Suppose one condensate, say  condensate 2, 
has a particle number much smaller
than that of condensate 1, but still large enough to have a well-defined
interface and surface. 
Then condensate 2 may form a droplet inside 
condensate 1 in a phase segregated regime.
Because of the finite surface tension, this droplet may 
move to and stay at the edge of the trap 
to reduce the common boundary  length in 2D (or area in 3D) 
between condensates 1 and 2, to minimize the surface energy.
This tendency may be called the floating of the condensate 2 droplet.

\subsection{ Finite trap size effect II: lower critical
     condensate-atom number for phase segregation}

In the above analysis, we have implicitly assumed that the thickness of the 
interface is much smaller than the trap size, such that we have 
a well-developed phase segregation.
We examine the limitation of this assumption here.
According to the above analysis, the penetration of  condensate 2 into
1 is determined by $\Lambda$, and the recovery of  condensate 1 from the
presence of the interface  by $\xi$ or $\Lambda$ in the strongly 
or weakly segregated phase.
As an estimation, we may have the interface thickness $l_{12}$ as
\be
   l_{12} \sim  \xi  + \Lambda   \; . 
\ee
This implies that 
$l_{12} \propto {1}/{\sqrt{\rho_{10}} }$.
When the interface thickness is larger than the trap size,
$
   l_{12} > R
$,
we do not have a well defined phase segregated state.
A conclusion inferred from this is that, 
if the condensate atom number is too small, 
there is no complete phase segregation in the trap
even with a strong mutual repulsive interaction between the condensates.
 
We should point out that the analyzes in Secs. 2.4 and 2.5 
are based on the assumption that the two condensates have the 
same sets of healing length and penetration  depth.
This implies that the surface energy contributions caused by the 
edge of the square well trap are the same for both condensates.
Hence we have ignored their effect in the determination  of the symmetry
of the ground state.
In general, this may not be true, 
and one has to consider the competition among
four lengths: two healing lengths and two penetration depths, as well as
the surface energy from the trap edge.
This will generate an even richer physics than what has been presented 
above. 
We believe our above discussions have provided the framework
for further explorations. 
One such example will be discussed in Sec. 3.

\section{ Numerics and discussions }

In terms of the atomic scattering lengths of condensate atoms
$a_{ii}$, the interactions are 
$G_{ii} = {4\pi \hbar^2 a_{ii} }/{m_i } $.
The typical value of $a_{ii}$ for $^{87}$Rb is 
about $50$ \AA.
The typical density realized for the binary BEC mixture is 
about $\rho_{i0} \sim 10^{14}/$cm$^3$.
Hence the healing length is 
$\xi = \sqrt{  ({\hbar^2}/{2m}) ({1}/{G_{ii} \rho_{i0} } ) }
     = \sqrt{1/(8\pi a_{ii}\rho_{i0} ) } \sim 3000$ \AA.
For the different hyperfine states of $^{87}$Rb,
we may take 
${G_{12} }/{ \sqrt{ G_{11}G_{22} } } = 1.04$,
whose precise value is uncertain and may be smaller.
Then the penetration depth is 
$\Lambda = \xi/\sqrt{ {G_{12} }/{ \sqrt{ G_{11}G_{22} } } - 1 }  
     \sim 1.5 \mu$m.
The length scale for  the harmonic trap potential determined by oscillator
frequencies
ranges from 1.3 to 4 $\mu$m, 
and the condensate occupies a region with 
a diameter of about 20 $\mu$m for about $10^6$ atoms,
which is comparable to or larger than the size of the interface thickness.
Hence it is reasonable in practice 
to apply the mean-field results obtained for a square well potential
to the case of a harmonic trapping potential, because 
on the scale of interface thickness the trapping potential appears smooth.
The measurement of the penetration depth 
can be used to determine 
the mutual repulsive interaction in the $^{87}$Rb system, 
which is in the weakly segregated phase, because $G_{12}$ is
believed \cite{jila} to be slightly larger than $\sqrt{G_{11} G_{22} }$.

Now we consider whether or not the ground state symmetry of the hyperfine
states $^{87}$Rb can be broken.
As pointed out at the end of Sec. 2.5, there is a complication arising 
from the trap edge surface energy contribution, due to the
difference in healing lengths or interactions.
In accordance with the experimental situation and for the sake of 
simplicity of analysis, 
we take two condensates having an equal number of particles. 
The differences in interactions are small: 
$\{ \delta_i = (G_{ii} - G_{12} )/G_{12} , \,  i=1,2\}$   
are close to zero,
 and their sum $\delta_1 + \delta_2 \sim 0$.
A condensate near the edge of the square well potential is 
identical to the strong segregated phase, because there is
no penetration into the hard wall.
The surface tension at the trap edge
can be readily evaluated according to an expression similar to 
Eq. (15) in the strongly segregated phase but with only one condensate.
The difference in those surface tensions is then, up to the first order in
$\delta_i$,
\[
  \Delta \sigma =  \sigma_1 -\sigma_2 
         = \frac{1}{2\sqrt{2} } \xi_1 G_{11} \rho_{10}^2 
                   \times \frac{1}{4} (\delta_2 - \delta_1 )  \; .
\]
Here we have used Eq. (5) that $\sqrt{G_{22} } \, \rho_{20} = 
\sqrt{G_{11} } \, \rho_{10} $. 
If there were no symmetry breaking in the ground state, 
the condensate with the lower surface energy at the trap edge, 
say condensate 2, 
will stay close to the trap edge. Condensate 1 stays inside.
If there is a symmetry breaking such that each occupies half of the
trap as discussed in Sec. 2.4, 
then condensate 1 will get in contact with half of the trap edge.  
There is an increase in total energy as
\be
  \Delta E_{edge} = \pi R \Delta \sigma  =
          \frac{\pi }{8\sqrt{2} }  R  \, \xi_1 G_{11} \rho_{10}^2 
                   \times (\delta_2 - \delta_1 )  \; .   
\ee
For the symmetry breaking to occur, this surface energy cost from the 
trap edge must be smaller than 
the interface energy gained, which is
\be
   \Delta E_{interface} = E_s - E_b = \frac{ \pi - \sqrt{2} }
    { 2 }  R \, \xi_1 G_{11} \rho_{10}^2 
           \times (\delta_2 - \delta_1) \; .
\ee
Here we have used 
 $\xi_i/\Lambda_i = (\sqrt{2} / 4) \; (\delta_2 - \delta_1 ) $ and
 $\delta_1 + \delta_2 =0$.
One can readily check now that indeed 
$\Delta E_{interface} > \Delta  E_{edge}$.
We note that in the case of the harmonic trapping potential,
the density of a condensate is smaller near the edge, which gives a even
smaller edge surface energy contribution.
Therefore the symmetry breaking will occur in the system of 
hyperfine states of $^{87}$Rb according to the present analysis.
We believe this is precisely what was observed 
in a very recent experiment \cite{further}. 
Furthermore, since it is in the weakly segregated phase, one
should also expect that the symmetry break 
occurs only for the density of  each individual condensate, 
not for the total density.
Again, this is what was reported in Ref. \\onlinecite{further}.

Given the size of the trap to be on the order of 20 $\mu$m,
the interface thickness must be smaller than this length to have a 
well-defined segregated phase.
For the different hyperfine states of $^{87}$Rb, this implies that, 
according to Eq. (18), a lower critical number $N_c$ of condensate atoms
$N_C =  {1}/{ ( {G_{12} }{\sqrt{G_{11} G_{22} } } - 1 ) } 
        \; { V^{1/3} }/{ (8\pi a_{ii} ) } 
    \sim 4000 $.
We should point out that the precise value of the lower critical 
atom number depends on details of a realistic trapping potential, such 
as the oscillator frequency and the anisotropy ratio. 
For a condensate-atom number of less than this value, 
there is no complete phase segregation.

\section{Conclusion}

From the mean-field analysis, 
by tuning the strength of the mutual repulsive 
interaction we have found that there are two segregated phases:
The interface profile 
is determined by the penetration depth in the weakly segregated phase , 
and by the healing length in the strongly segregated phase. 
The broken cylindrical symmetry state starts to appear in the weakly
segregated phase for each condensate, and persists into the strongly 
segregated phase.
For the total condensate density, the cylindrical symmetry 
is maintained in the weakly segregated phase, and disappears in 
the strongly segregated phase.
We have also found that 
a condensate droplet inside another condensate in a
segregated phase tends to move to and to stay at the trap boundary, 
and that 
if the condensate atom number is smaller than 
a critical value, there is no well-developed phase segregation.
A comparison between the present results and a recent experiment
has suggested the weakly segregated phase has been observed in $^{87}$Rb
condensates.

{\ }

\noindent
{ This work was supported in part  by a grant from NASA (NAG8-1427).  }

{\ }

\noindent
$^{\star}$Permanent address: Department of Theoretical Physics,
Ume\aa{\ }University, 901 87, Ume\aa, Sweden


\begin{thebibliography}{99}

\bibitem{ho}
   T.L. Ho and V.B. Shenoy, Phys. Rev. Lett. {\bf 77}, 3276 (1996).
\bibitem{jila}  
  C.J. Myatt, E.A. Burt, R.W. Ghrist, E.A. Cornell, and C.E. Wieman, 
     Phys. Rev. Lett. {\bf 78}, 586 (1997).
\bibitem{further}
%
 D.S. Hall, M.R.M. Matthews, J.R. Ensher, C.E. Wieman, and E.A. Cornell, 
    cond-mat/9804138. 
\bibitem{molmer}
  K. M\o lmer, Phys. Rev. Lett. {\bf 80}, 1804 (1998).
\bibitem{theoretical}
  E.V. Goldstein and P. Meystre, Phys. Rev. {\bf A55}, 2935 (1997);

  B.D. Esry, C.H. Greene, J.P. Burke, and J.L. Bohn,
    Phys. Rev. Lett. {\bf 78}, 3594 (1997);

  P. \"Ohberg and S. Stenholm, Phys. Rev. {\bf A57}, 1272 (1998);
 
  H. Pu and N.P. Bigelow, Phys. Rev. Lett. {\bf 80}, 1134 (1998);
 
  D. Gordon and C.M. Savage, cond-mat/9802247.
\bibitem{law}
   C.K. Law, H. Pu, N.P. Bigelow, and J.H. Eberly,
    Phys. Rev. Lett. {\bf 79}, 3105 (1997).
\bibitem{cta}
  S.T. Chui, B. Tanatar, and P. Ao,  Phys. Rev. {\bf A }(in print).
\bibitem{nlse}
   O. Penrose,  Phil. Mag. {\bf 42}, 1373 (1951);
  
   E.P. Gross, Nuovo Cimento {\bf 20}, 454 (1961);

   L.P. Pitaevskii, Sov. Phys. JETP {\bf 13}, 451 (1961);
  
   E. Demircan, P. Ao, and Q. Niu, Phys. Rev. {\bf B54}, 10027 (1996).
 
\end{thebibliography}
\end{document}